\documentclass[aps]{revtex4}
\usepackage{graphicx}
\usepackage{bm}
\usepackage{amsmath}
\usepackage[T1]{fontenc}
\usepackage{mathrsfs,ae,dsfont}

\begin{document}

\title{Confinement Admits Chiral Symmetry Breaking via Bag}

\author{Hai-Jun Wang}
 \email{hjwang@jlu.edu.cn}
\author{Hong Wang}
 \email{whong17@emails.jlu.edu.cn}
\affiliation{Center for Theoretical Physics and School of Physics,
Jilin University, Changchun 130012, China}

\begin{abstract}
In this paper, it is pointed out for the first time that the linear
effective potential between quarks is intrinsically relating to the bag
model while concerning the asymptotically-free nature of colours. Based on
the relationship we employ the symmetry method to analyze the
quark-anti-quark system. By imposing the Poincare invariance on the
quark-anti-quark bound-state, and translating the chiral transformation to
its spatial manifestation, we can infer why the chiral symmetry breaking
happens. Applying this knowledge to deep inelastic scattering we reach the
conclusion that the measured proton spin in scattering experiments should be
uncertain quantity.
\end{abstract}

\maketitle

keywords: {confinement; linear potential; bag model; chiral symmetry
breaking;conformal group;spin structure of hadrons }

\section{Introduction}

The non-perturbative part of QCD can hardly be resolved
analytically. Different methods ~\cite{Hooft81, Kugo79, Gribov99,
Parth10, Alko08, Siringo16, He13, Efimov02, Kondo15} have been
proposed to simplify the process of searching solutions, all of
which conclude with a linear potential between a quark and an
anti-quark (It is also known as the flux tube.). In addition,
Regge trajectories can be viewed as a compelling evidence for the
rationale of linear potential. The linear potential has shown its
essence in calculating the spectra of mesons with heavy quarks and
has produced reliable results in agreement with experiments
~\cite{Jia17, Sharov13, Afonin16, Afonin14, Lewis94, Roy13}. Up to
date it has been believed that the linear potential is not only
compatible with colour confinement, but also a common
interpretation of colour confinement.

~~\\

Chiral symmetry breaking ($\chi $SB)~\cite{Ecker95, Plant98,
Chang11, Alko08, Siringo16, Glozman12, Shuryak14, Pena17} is the
other side of non-perturbative QCD. Reconciling colour confinement
with $\chi $SB becomes critical for understanding the
non-perturbation of QCD~\cite{Alko08, Glozman12, Siringo16,
Roberts17, Suganuma17}. In the classic papers~\cite{Casher 79,
Casher 80}, Banks and Casher have analyzed the relationship
between $\chi $SB and confinement, where they obtained the
conclusion that fermion-anti-fermion bound state definitely leads
to $\chi $SB. The analysis exposed the critical role of the
spin-related interaction, so the Authors introduced a parameter
relating to the ratio of spin contribution and
orbital-angular-momentum contribution. Thereafter people also find
other ways to infer $\chi $SB ~\cite{Bicudo04, Syna08, Mall12,
Glozman12, Alko08, Maris03, Kumar04}. With the accumulation of the
experimental data of angular momentum of hadrons~\cite{Ji13}, it
is time for us to revisit the relationship among linear potential,
$\chi $SB, and the angular momentum.

~~\\

\section{The relationship between linear potential and bag}

Here we will associate linear potential with $\chi $SB via bag~\cite{Jaffe74}%
. Physically, there exists a true bag in hadron, not just a model any
longer, if only we combine the sea gluons with a quark. As for the linear
potential, it provides a constant force $\sigma $, which implies in its own
right that the lines of colour force (henceforth named ''colour lines'')
sent from one quark are all absorbed by another (anti-)quark. In terms of
gluon field, that means the gluons sent from one quark are all absorbed by
another (anti-)quark, without any leaking. Moreover, knowing that protons
always dress pion cloud, similarly we regard the gluons being dressed by the
quark/anti-quark (to be the quark or the anti-quark depending on the
reference frame we choose), subsequently we recognize that the dressing
quark/anti-quark becomes spatially extended. Without losing generality, we
propose that it is the quark that dresses all of the virtue gluons, forming
a closed bag surrounding the anti-quark. And the anti-quark remains to be a
point particle. But why should it be bag?

~~\\

The quark extended in space will be identical to either a point, or a bag
topologically, according to the following arguments. Holding the picture
that strong interaction works certainly by exchanging boson particles
between points, then it is the familiar scenario whence the force is
reversely proportional to an imagined spherical area $4\,\pi \,r^2$, where $%
r $ is the distance between the two points. If only the quark is
not closed bag, this scenario holds, coinciding with asymptotic
freedom~\cite {Wang03,Gross73}. Now assuming that the
spatially-extended quark forms a closed bag and, the interaction
between a point in quark and the anti-quark is still proportional
to $1/r$ [caution that the quark is sizable and the anti-quark
remains a point particle], then the interaction term between the
quark and anti-quark in Schrodinger equation should be an integral
form like $\int_{\text{bag}}\frac 1r\psi (\vec r)\,d\vec
r=\int_{\text{bag}}\frac 1r\psi (\vec r)\,r^2drd\Omega =4\pi
\int_{\text{bag}}r\psi (\vec r)\,dr=4\,\pi \,R\psi (\vec R)dR_0$.
The last step assumes the ideal case when the quark(bag) forms a
spherical surface with fixed thickness $dR_0$. So far one could
conclude qualitatively with the linear potential.

~~\\
\begin{table}[tbp]
\begin{ruledtabular}
\begin{tabular}{cccccccc}
 States $(b\bar b)$ & $V_1(r)$ &$V_2(r)$ & Experiment(MeV) \cite {Fulcher94}\\
\hline 1S& 9464 & 9464 & 9460\\
2S& 11169 & 9572 & 10023 \\
3S& 13982 & 9660 & 10355\\
4S& 17862 & 9738 & 10579\\
1P& 9459 & 9527 & 9859\\
2P& 11158 & 9621 & 10232\\
3P& 13966 & 9702 & 10512\\
4P& 17842 & 9776 & \\
\end{tabular}
\end{ruledtabular}
\caption{We use this table to illustrate the almost equivalence between $%
[1/r]+bag$ and the linear potential model, where $V_1(r)=-\frac{\alpha _s}r$
while $0<r<R$ and $V_1(r)$ becomes $\infty $ while $r>R$, $V_2(r)=\sigma \,r$
as usual defined. The parameters are $m_b=4660$\thinspace MeV, $\sigma
=3.3\times 10^4\,$MeV$^2$, $\alpha _s=0.356$, $R=1.925$fm. Please note that
the bag here is thin without thickness.}
\label{tab:table 1}
\end{table}
~~\\


We have used the potential $[1/r]+bag$ to calculate the spectra
for some states and the results are compared with those using
potential $[\sigma \,r]$ only. One notes that there does exist the
set of parameters for the two methods to lead to the same tendency
and similar separation between energy levels [Table I.]. Now the
bag looks indispensable in consistently bridging the asymptotic
potential $[1/r]$ and the confining potential $[\sigma \,r]$,
which are combined to be the very useful Cornell Potential ~\cite
{Trev13,Lewis94}. Surprisingly the total effect of an asymptotic
potential plus bag is almost equivalent to a linear potential
$[1/r]+bag\sim [\sigma \,r]$. The bag plays a role in associating
the perturbative dynamics with nonperturbative dynamics. The bag
also manifests non-Abelian characteristic since it cannot be
penetrated by other colour lines (gluons) which are all absorbed
by the dressed gluons of the quark, i.e. the
three-line-interaction-vertex of gluons works. It is totally a
non-Abelian property not pertaining to the dressed photons by
electron. $[1/r]+bag$ is in some sense better than $[\sigma \,r]$
since $[\sigma \,r]$ has its well-known Abelian
origination~\cite{Hooft81, WangQ2000}. Now based on the above
arguments we accept the effectiveness that one of the quark and/or
the anti-quark forms bag in the extreme state, i.e. meson ground
state. We can picture a process for the above argument: with the
decrease of system's energy, one end of the flux tube (quark side)
swells to be like an umbrella then finally forms the bag. In the
process the gluon field evolves to be like many extra ribs to
support the umbrella. In the final bag state the Regge
trajectories remain effective in evaluating hadron spectra. If
only the colour lines are not leaked, then the linear potential
holds, as well as the effective flux tube.

~~\\

There already existed demonstration on how the bag induces $\chi
$SB from Lagrangian~\cite{Bron09}. But the explanation doesn't
comply physically with the massless situation while the chiral
symmetry is restored, since it implies that the bag would
disappear for pions or any goldstone bosons. And we are tempted to
employ symmetry methods to reconcile the conflict.

\section{The realization of chiral transformation}

In the non-perturbative regime, where the bag (with thickness $dR_0$)
locates, the smallest symmetric group is assumed to be Poincare group which
includes Lorentz group as sub-group. Of course, the quark-anti-quark system
should be represented by a Lagrangian as described above,
\begin{equation}
\mathscr{L}=\mathscr{L}(\text{asymptotic potential}+\text{classical bag})_{%
\text{Poincare invariance}},  \label{Lag}
\end{equation}
it now represents a two-quark system. In previous paragraphs we have used
the sign $[1/r]$ to represent asymptotic potential. In our context of
discussion, we don't need to know the concrete form of the Lagrangian. We
will only study how chiral symmetry affects the Poincare invariance,
reflected by the effect of chiral transformation on Dirac spinors and by its
commutation relations with generators of Poincare group.

As for the Lorentz transformation affecting Dirac spinor, it is a
sort of effect rotating the spinors within the configuration
spanned by different z-component spin value, which corresponds
partially to rotating a reference frame in true 3-D space.
Moreover, besides the 3-D rotation, the ''rotation'' also includes
the ''transformation'' between different boost frames. The above
two sorts of rotations don't interchange spinors belonging
separately to positive energy and negative energy. However, the
chiral transformation does interchange, we will recognize that in
what follows.

Since the chiral transformation shares the same generator $\gamma ^5$ with
scaling transformation (one of the generators of conformal group) in spinor
representation, differing only in an imaginary number $i$, it is necessary
to introduce the conformal group first. The 4-dimension Conformal Group~\cite
{Budi79,Yu13,Budi79A, Mack69}, with Poincare group as its sub-group, its
generators in differential form are\cite{Budi79,Ma99},
\begin{eqnarray}
D &=&\,x_\mu \frac \partial {\partial x_\mu }\text{, }M_{\mu \nu }=i(x_\mu
\frac \partial {\partial x^\nu }-x_\nu \frac \partial {\partial x^\mu })%
\text{,}  \nonumber \\
P_\mu &=&i\frac \partial {\partial x^\mu }\text{ , }K_\mu =-i(x^2\frac
\partial {\partial x^\mu }-2x_\mu x^\nu \frac \partial {\partial x^\nu })%
\text{,}  \label{ccc}
\end{eqnarray}
the following commutation relations hold,
\begin{eqnarray}
\lbrack M_{\mu \nu },M_{\rho \sigma }] &=&i(g_{\nu \rho }M_{\mu \sigma
}+g_{\mu \sigma }M_{\nu \rho }-g_{\mu \rho }M_{\nu \sigma }-g_{\nu \sigma
}M_{\mu \rho }),  \nonumber \\
\lbrack M_{\mu \nu },P_\rho ] &=&i(g_{\nu \rho }P_\mu -g_{\mu \rho }P_\nu ),
\nonumber \\
\lbrack D,P_\mu ] &=&-P_\mu \text{, }[D,K_\mu ]=K_\mu ,  \nonumber \\
\lbrack D,M_{\mu \nu }] &=&0  \nonumber \\
\ \ [M_{\mu \nu },K_\rho ] &=&i(g_{\nu \rho }K_\mu -g_{\mu \rho }K_\nu )
\label{dd} \\
\ [P_\mu ,K_\rho ] &=&-2\,i\,(g_{\mu \rho }\,D+M_{\mu \rho })  \nonumber
\end{eqnarray}
And from the original definition of conformal group~\cite{Cartan37}, the
spinor representation can be derived ~\cite{Han15},
\begin{eqnarray}
\gamma _i\gamma _j &\longrightarrow &i(x_j\frac \partial {\partial
x^k}-x_k\frac \partial {\partial x^j})\longrightarrow M_{jk}  \nonumber \\
\gamma _0\gamma _i &\longrightarrow &i(x_i\frac \partial {\partial
x^0}-x_0\frac \partial {\partial x^i})\longrightarrow M_{0k}  \nonumber \\
\gamma _5 &\longrightarrow &\,x_\mu \frac \partial {\partial x_\mu
}\longrightarrow D  \nonumber \\
\gamma _\mu (1+\gamma _5) &\longrightarrow &i\frac \partial {\partial x^\mu
}\longrightarrow P_\mu  \nonumber \\
\gamma _\mu (1-\gamma _5) &\longrightarrow &-i(\frac 12x_\nu x^\nu \frac
\partial {\partial x_\mu }-x_\mu x_\nu \frac \partial {\partial x_\nu })%
\text{ }\longrightarrow K_\mu \text{.}  \label{mapping}
\end{eqnarray}
where $\gamma _\mu $'s are Dirac $\gamma -$matrices and we use $%
\longrightarrow $ to represent the accurate mappings, and the same
commutations as eq.(\ref{dd}) can be examined. We have recognized that the
role of operator $\mu \frac{\text{d}}{\text{d}\mu }$ (or $x_\mu \frac
\partial {\partial x_\mu }$) in the conformal group is equivalent to that of
the scaling operator $D$, with its spinor representation being $\gamma _5$.~

By definition the scaling part of the conformal group will either stretch or
press the whole bag homogenously while varying the scale. It needs energy
injection from or leakage to outside \cite{Han15}. Now let's have a glimpse
at how scaling transformation and chiral transformation change the spinors
of free particle. The chiral transformation $e^{i\frac u2\gamma _5}$and the
scaling transformation $e^{\frac u2\gamma _5}$ differ in an imaginary phase
factor $i $. Now let the transformations perform on the simplest spinor $%
\left(
\begin{array}{c}
1 \\
0 \\
0 \\
0
\end{array}
\right) $, which are
\begin{equation}
e^{\frac u2\gamma _5}\left(
\begin{array}{c}
1 \\
0 \\
0 \\
0
\end{array}
\right) =\left(
\begin{array}{c}
\cosh \frac u2 \\
0 \\
\sinh \frac u2 \\
0
\end{array}
\right) \text{ ,}  \label{A2}
\end{equation}
and
\begin{equation}
e^{i\frac u2\gamma _5}\left(
\begin{array}{c}
1 \\
0 \\
0 \\
0
\end{array}
\right) =\left(
\begin{array}{c}
\cos \frac u2 \\
0 \\
i\sin \frac u2 \\
0
\end{array}
\right) \text{ .}  \label{A3}
\end{equation}
It is noted that they separately satisfy normalizations ${\psi }^{\dagger }{(%
}x{)\gamma }_0{\psi }{(}x)=u_1^{*}u_1+u_2^{*}u_2-u_3^{*}u_3-u_4^{*}u_4=1$,
and ${\psi }^{\dagger }{(}x{)\psi }{(}%
x)=u_1^{*}u_1+u_2^{*}u_2+u_3^{*}u_3+u_4^{*}u_4=1$ , where ${\psi }{(}x{)}$
is a complex spinor, and ${\psi }^{\dagger
}=(u_1^{*},u_2^{*},u_3^{*},u_4^{*})$, e.g. the eq.(\ref{A3}) is ${\psi }%
^{\dagger }=(\cos \frac u2,0,-i\sin \frac u2,0)$. Though the former is
Minkowski and the latter is Euclidean, the results keep the same
\[
\cosh ^2\frac u2-\sinh ^2\frac u2=\cos ^2\frac u2+(-i\sin \frac u2)(i\sin
\frac u2)=1\text{.}
\]

Now let's examine the effect of chiral transformation on spinor in more
detail. A regular Dirac spinor can be written as
\[
A\left(
\begin{array}{c}
s_1 \\
s_2
\end{array}
\right) _{4\times 2}\varphi _\alpha
\]
where $A$ is a normalization constant, and $\varphi _\alpha $, $\alpha =1,2$
are $2\times 1$ spin eigen states or helicity eigen states as $\left(
\begin{array}{c}
1 \\
0
\end{array}
\right) $ or $\left(
\begin{array}{c}
0 \\
1
\end{array}
\right) .$ The part $\left(
\begin{array}{c}
s_1 \\
s_2
\end{array}
\right) $ can be written as $\left(
\begin{array}{c}
1 \\
\frac{\vec \sigma \cdot \vec p}{E+m}
\end{array}
\right) $ or $\left(
\begin{array}{c}
\frac{\vec \sigma \cdot \vec p}{E+m} \\
1
\end{array}
\right) $ corresponding to free particle or free anti-particle respectively.
Now let's check what if $e^{i\frac u2\gamma _5}$ performs on $\left(
\begin{array}{c}
s_1 \\
s_2
\end{array}
\right) $.
\[
e^{i\frac u2\gamma _5}\left(
\begin{array}{c}
s_1 \\
s_2
\end{array}
\right) =\cos \frac u2\left(
\begin{array}{c}
s_1 \\
s_2
\end{array}
\right) +i\sin \frac u2\left(
\begin{array}{c}
s_2 \\
s_1
\end{array}
\right) \text{ .}
\]
We note this is another sort of spinors rotation, which mixes the states of
both the particle and the anti-particle.

Then let's consider here the spatial effect of chiral
transformation on the quark-anti-quark system. According to the
above analysis, in such case the chiral transformation is somehow
equivalent to the partial exchange of quark state and anti-quark
state, another sort of abstract ''rotation''. With the
aforementioned bag picture, if the quark and anti-quark masses are
other than zero, one notes that the center of mass changes from
anti-quark to quark while performing chiral transformation
$e^{i\frac \pi 2\gamma _5}$ [Fig 1]. The chiral transformation
becomes ''observable'' due to the need of injecting energy to
displace the center of mass. Accordingly while the quark and
anti-quark are massless, it is clear now such chiral
transformation is not observable, i.e. the chiral symmetry holds.
This is the picture of $\chi $SB in spatial manifestation via bag.
Now $\chi $SB can be inferred from the confinement.

~~\\

\begin{figure}[tbp]
\includegraphics[width=12cm,height=16cm,angle=270]{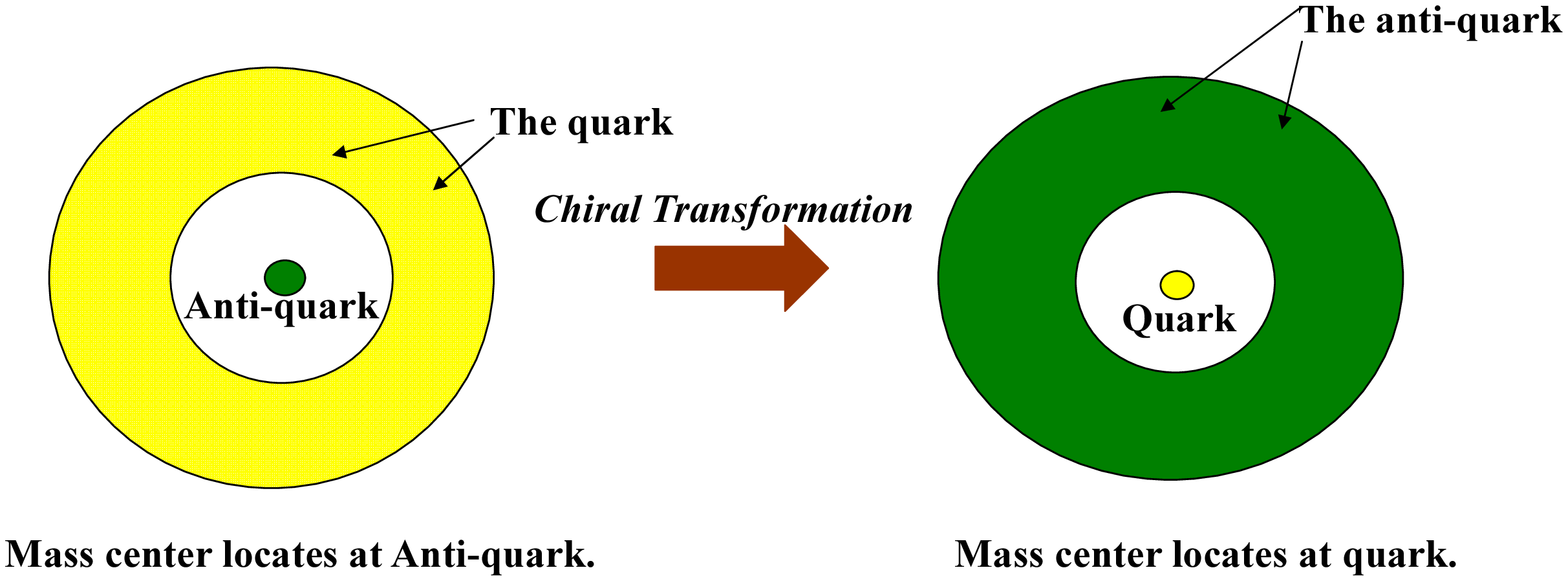}
\caption{The two-quark system before and after the chiral transformation.}
\label{fig:art}
\end{figure}


~~\\

To put the above analysis alternatively, one notes the Lagrangian eq.(\ref
{Lag}) is invariant under Poincare group, which includes translations of
energy and momentum. According to the eq.(\ref{mapping}) the chiral
transformation happens not to commutate with these four translations, [$%
\gamma _5,\gamma _\mu (1+\gamma _5)]\neq 0$, especially the energy part.
Thus while the energy (mass) of the system changing (displacement of mass
center), then the chiral symmetry becomes varying, i.e. breaking or
restoring.

~~\\

\section{Discussions}

The model in this paper offers a new insight into the spin structure of
hadrons ~\cite
{Leader,FWang1,FWang2,Ji12,Ji97A,Ji97B,Ma,Jaff95,Bass05,Waka10,Waka11,Hatta11}
. In contrast to confinement, while a large amount of external energy is
injected into hadrons, we know the asymptotic freedom would be reached by
scattered fragments (including partons, quarks etc.) and chiral symmetry
will be restored by these fragments. The fragments take almost zero masses.
In such ultra-high-energy scattering case, the aforementioned pictured
process evolves backward with the increase of energy, i.e. from the bag back
to flux tube. Actually there are many such flux tubes in the fragment jets~%
\cite{ExP of Gluon,Alex17}. Now these flux tubes are surrounded almost
isotropically by identical particles (fermions like partons, quarks or
bosons like meson). The flux tubes together with identical particles form
dual composite particles, whose angular momenta exclusively depend on
environment ~\cite{Wilczek82}.

We call these composite particles anyons, they live in 2-dimension
environment. How can this 2-dimension condition be satisfied in
such scattering case? Without losing generality, we know the
composite particles must locate in
one of the final-state-planes, and in fact only one plane dominates there ~%
\cite{Collider}. As for how flux tubes are decomposed to those
perpendicular to the plane and those parallel to the plane, please
refer to ~ \cite {Perpendicular}. The interactions among composite
particles are complicated according to ~\cite{Wilczek82}, as well
as the statistics and the total angular momenta of these
particles. So in this hot state of mixed particles, the angular
momentum is consequently not certain due to the anyons.
Accordingly, the sum of angular momenta we measured from the jets
cannot be used to infer reversely the original spin of hadrons.
Alternatively, maybe from the factor $\frac e{\delta m}$ we can
also conclude that the total spin of quarks are uncertain, which
causes the total angular momenta uncertain. This uncertainty stems
from the 2-d condition and gauge transformation (the
transformation of gauge potential $\vec A$)~\cite{Wilczek82}, plus
the varying chiral parameter (the quark mass). The result
coincides with the reference~\cite{Ji13}, telling that only while
quarks are in light-cone gauge, could the spin and angular
momentum be physical observables.

Another interesting aspect of this research is the topology shown
by transformation of the chiral symmetry, i.e. while interchanging
the spatial position of quark and anti-quark. It is very mimicking
the Mobius band in four dimension, the Klein bottle~\cite{Klein}.
We may call it the inner-outer transform, which sounds a bit
strange since one cannot imagine how to drag the anti-quark from
the inner of the closed bag to its exterior, unless you think of
things like Klein bottle realized in 4-dimension space. Here we
have attributed the topology effect to single quark. The
alternative way of expressing such topology is to use the $\chi
$SB result directly, leading to such as PCAC or the linear sigma
model, from which one can derive topology effects for emerging
quasi-particles ~\cite{MaBook18}. At the present stage we
conjecture that these two ways of deriving topology may be
equivalent. To speak alternatively, to view the quark being
extended in space, is somehow equivalent to understand hadrons as
something emergent formed by a lot of point particles.
Realistically, while the energy of quark is truly low and thus its
wavelength becomes very large, the quark becomes spatially
extended completely, with dressed gluons as a whole entity. It
could be a constituent quark.

~~\\

\begin{acknowledgments}
H. J. W. is grateful to Prof. Duojie Jia (Northwest Normal Univ.,
Lanzhou), Prof. Yu-Xin Liu (Peking Univ, Beijing), Prof. Qing Wang
(Tsinghua U., Beijing), Prof. Yong-Liang Ma (Jilin Univ.), Prof.
Mannque Rho (IPhT, Saclay) for their substantial and heuristic
discussions. The work was supported in part by National Science
Foundation of China (NSFC) under Grant No. 11647304, 11475071, and
11547308.
\end{acknowledgments}

~~\\


\begin{thebibliography}{99}
\bibitem{Hooft81}  G. t' Hooft, Nucl. Phys. B 190, 455 (1981).

\bibitem{Kugo79}  Taichiro Kugo and Izumi 0jima, Supplement of the Progress
of Theoretical Physics, No. 66, 1979.

\bibitem{Gribov99}  V. N. Gribov, Eur.Phys.J. C 10 (1999) 91-105.

\bibitem{Parth10}  R. Parthasarathy, arXiv:1003.1209v1.

\bibitem{Alko08}  Reinhard Alkofer, Christian S. Fischer, Felipe J.
Llanes-Estrada et al, Ann. phys 324 (2009)106-172. or arXiv:0804.3042v1.

\bibitem{Siringo16}  Fabio Siringo, Phys. Rev. D 94, 114036 (2016).

\bibitem{He13}  Han-Xin He and Yu-Xin Liu, arXiv 1307.4485v3.

\bibitem{Efimov02}  G. V. Efimov and G. Ganbold, Phys. Rev. D65 054012(2002).

\bibitem{Kondo15}  Kei-Ichi Kondo, Seikou Kato, Akihira Shibata, Toru
Shinohara, Phys. Rep. 579 (2015)1-226.

\bibitem{Afonin16}  S.S. Afonin and I. V. Pusenkov, arXiv:1606.05218v1.

\bibitem{Afonin14}  S.S. Afonin and I. V. Pusenkov, Phys. Rev. D 90, 094020
(2014).

\bibitem{Lewis94}  Lewis P. Fulcher, the spinless Salpeter equation and the
Cornell potential, Phys. Rev. D 50, 447 (1994).

\bibitem{Roy13}  Sabyasachi Roy and D K Choudhury, Phys. Scr 87 (2013)065101.

\bibitem{Jia17}  Duojie Jia, Cheng-Qun Pang, and Atsushi Hosaka, Int. J.
Mod. Phys. A 32, 1750153 (2017).

\bibitem{Sharov13}  G. Sharov, String Models, Stability and Regge
Trajectories for Hadron States, arXiv:1305.3985.

\bibitem{Ecker95}  Gerhard Ecker, Chiral perturbation theory, Prog. Part.
Nucl. Phys. 35 (1995), pp. 1C80.

\bibitem{Glozman12}  L. Ya. Glozman, Confinement, arXiv:1211.7267v1.

\bibitem{Shuryak14}  Edward Shuryak, Nucl. Phys. A 928, 138 (2014).

\bibitem{Pena17}  M. T. Pena, Elmar P. Biernat, Alfred Stadler, Confinement
and Chiral-Symmetry Breaking in the Covariant Spectator Theory, Few-Body
Syst, DOI 10.1007/s00601-015-0955-2.

\bibitem{Plant98}  Robert S. Plant and Michael C. Birse, Nucl.Phys. A628
(1998) 607-644.

\bibitem{Chang11}  Lei Chang, Yu-Xin Liu, Craig D. Roberts,
Phys.Rev.Lett.106:072001 (2011).

\bibitem{Roberts17}  Craig D. Roberts, Perspective on the Origin of Hadron
Masses, Few-Body Syst, DOI 10.1007/s00601-016-1168-z.

\bibitem{Suganuma17}  Hideo Suganuma, Takahiro M. Doi, Krzysztof Redlich,
and Chihiro Sasaki, Some relations for quark confinement and chiral symmetry
breaking in QCD, EPJ Web of conferences 137, 04003 (2017), XII th Quark
confinement and the Hadron specturm.

\bibitem{Casher 79}  A. Casher, Phys. Letts 83B, 395 (1979).

\bibitem{Casher 80}  T. Banks and A. Casher, Nucl. Phys. B169, 103 (1980).

\bibitem{Bicudo04}  P. Bicudo and G.M. Marques, Phys. Rev. D70 094047(2004).

\bibitem{Syna08}  Franziska Synatschke, Andreas Wipf, and Kurt Langfeld,
Phys. Rev. D 77, 114018(2008).

\bibitem{Mall12}  Elyse-Ann O¡¯Malley, Waseem Kamleh, Derek Leinweber, and
Peter Moran, Phys. Rev. D 86, 054503 (2012).

\bibitem{Maris03}  P. Maris, A. Raya, C.D. Roberts, and S.M. Schmidt, Eur.
Phys. J. A 18, 231(2003).

\bibitem{Kumar04}  Alok Kumar, R. Parthasarathy, Phys. Lett. B 595 (2004)
373.

\bibitem{Jaffe74}  A. Chodos, R. L. Jaffe, K. Johnson, and C. B. Thorn,
Phys. Rev. D 10(1974)2599.

\bibitem{Wang03}  Hai-Jun Wang, Hui Yang, and Jun-Chen Su, Phys. Rev. C 68,
055204 (2003).

\bibitem{Gross73}  David J. Gross and Frank Wilczek, Phys. Rev. Lett. 30,
1343(1973).

\bibitem{Fulcher94}  Lewis P. Fulcher, Phys. Rev. D 50 (1994)447.

\bibitem{Trev13}  L. A. Trevisan, Carlos Mirez, F. M. Andrade, Few-Body Syst
55:1055 (2013).

\bibitem{WangQ2000}  Qing Wang, Yu-Ping Kuang, Xue-Lei Wang, Ming Xiao,
Phys.Rev. D61 (2000) 054011.

\bibitem{Bron09}  Wojciech Broniowski, Quark models in physics of strong
interactions: from the MIT bag to chiral symmetry, Report number: Historical
overview, Coimbra, April 2009.$https:// cft.fis.uc.pt/Documents/qm_wb.pdf$.

\bibitem{Budi79}  P. Budini, Czechoslovak Journal of Physics B 29, 6 (1979).

\bibitem{Budi79A}  P. Budini, P. Furlan, R. Raczka, IL Nuovo Cimento A 52,
191 (21 Luglio 1979).

\bibitem{Yu13}  Yu Nakayama, arXiv:1302.0884 [hep-th].

\bibitem{Mack69}  G. Mack and Abdus Salam, Ann. Phys. 53, 174(1969).

\bibitem{Ma99}  Yufen Liu, Zhongqi Ma, Boyuan Hou , Commun. Theor. Phys.31,
481(1999).

\bibitem{Cartan37}  \'Elie Cartan, \emph{The Theory of Spinors}, Dover
Publications, Inc. 1981. This is a republication of the first version
published by Hermann, Paris. 1966.

\bibitem{Han15}  Lei Han, Hai-Jun Wang, Chinese Physics C 39, 093102 (2015).

\bibitem{Leader}  E. Leader, Phys. Rev. D 83, 096012 (2011).

\bibitem{FWang1}  Fan Wang, X.S. Chen, X.F. Lu, W.M. Sun, T. Goldman,
arXiv:0909.0798.

\bibitem{FWang2}  Xiang-Song Chen, Wei-Min Sun, Fan Wang, and T. Goldman,
Phys. Rev. D 83, 071901(R) (2011).

\bibitem{Ji12}  Xiangdong Ji, Xiaonu Xiong, and Feng Yuan, Phys. Rev. Lett.
109, 152005 (2012).

\bibitem{Ji97A}  Xiangdong Ji, Phys. Rev. D 55, 7114(1997).

\bibitem{Ji97B}  Xiangdong Ji, Phys. Rev. Lett. 78, 610 (1997).

\bibitem{Ma}  B. Q. Ma, J. Phys. G 17, L53 (1991); B. Q. Ma and Q. R. Zhang,
Z. Phys. C 58, 479 (1993).

\bibitem{Jaff95}  R. L. Jaffe, Phys. Today 48 (9), 24 (1995).

\bibitem{Bass05}  Steven D. Bass, Rev. Mod. Phys. 77, 1257C1302 (2005).

\bibitem{Waka10}  M. Wakamatsu, Phys. Rev. D 81, 114010 (2010).

\bibitem{Waka11}  M. Wakamatsu, Phys. Rev. D 83, 014012 (2011).

\bibitem{Hatta11}  Y. Hatta, Phys. Rev. D 84, 041701(R) (2011).

\bibitem{ExP of Gluon}  de Florian D, Sassot R, Stratmann M and Vogelsang W
2008 Phys. Rev. Lett. 101 072001 [28] Adare A et al 2009 (PHENIX
Collaboration) Phys. Rev. D 79 012003

\bibitem{Alex17}  C.Alexandrou, M. Constantinou, K. Hadjiyiannakou, et al,
Phys. Rev. Lett 119, 142002 (2017).

\bibitem{Wilczek82}  Wilczek F. Phys. Rev. Lett., 1982, 49: 957

\bibitem{Collider}  R. K. Ellis, W. J. Stirling, B. R. Webber, QCD and
Collider Physics, Cambridge Monographs on Particle Physics, Nuclear Physics
and Cosmology. 1996. pp. 63-75.

\bibitem{Perpendicular}  It is well known that there are jet-planes each
consisting of 4-momenta-conservation clusters in the scattering
experiments~\cite{Collider}. And certainly there are bars (flux
tubes) within a plane before the jet scattered away. For any bar,
we can do projection according to the amount of its energy and
momentum, projecting it both parallel to the plane and normal to
the plane. After doing this we get a lot of ''bars'' either
perpendicular to the plane or within (parallel to) the plane. For
the bars perpendicular to the plane, we view them as flux tubes.
As for the bars parallel to the plane, since the restoration of
chiral symmetry, the fermions staying at the both ends of the bar
become undistinguishable due to the $\chi $Symmetry restoration .
We view these fermions run around flux-tube. In such case the
flux-tubes and the identical fermions form the composites, which
were called anyons.

\bibitem{Ji13}  Xiangdong Ji, Jian-Hui Zhang, and Yong Zhao, Phys. Rev.
Lett. 111, 112002 (2013).

\bibitem{Klein}  https: $//en.wikipedia.org/wiki/Klein_bottle$

\bibitem{MaBook18}  Y. L. Ma and M. Rho, Effective Field Theories for Nuclei and Compact-Star Matter, World Scientific, Singapore, 2018. https://doi.org/10.1142/11072 .

\end{thebibliography}
\end{document}